# Flexible Cloud/User-Centric Entanglement and Photon Pair Distribution with Synthesizable Optical Router

Fabian Laudenbach, Bernhard Schrenk, Martin Achleitner,
Nemanja Vokić, Dinka Milovančev, and Hannes Hübel

*AIT Austrian Institute of Technology, Center for Digital Safety&Security / Security & Communication Technologies, 1210 Vienna, Austria.*
*Author e-mail address: bernhard.schrenk@ait.ac.at*

The practical roll-out of quantum communication technologies in optical networks and the adoption of novel quantum applications demand the distribution of single or entangled photons. Flexibility and dynamicity are paramount for the provision of quantum resources, in order to scale with the number of users and to meet the demand of complex network architectures. We present a quantum network architecture that features this degree of reconfigurability, without being restricted to a rigid physical-layer network based on purely passive multiplexing componentry. We leverage spectral assets of photon-pair sources, from the short-wavelength band to the L-band, and agile spatial switching at a remote optical network node, in order to realize a flexible distribution map that features different flavors, reaching from cloud-centric to user-centric quantum connectivity. Photon pair distribution is experimentally demonstrated between five users in a 17-km reach, tree-shaped optical network, with high visibility entanglement shared between three users of the network. Simultaneous distribution of photons to more than one user is enabled and the delivered photon rate can be dynamically adjusted. Penalty-free operation is confirmed for integrating a co-existing classical control channel within the quantum.

**I. Introduction**
Quantum entanglement, one of the best known and perplexing hallmarks of quantum mechanics is also an essential ingredient for many quantum information applications. In quantum communication in particular, entanglement provides a very useful resource. For example, quantum teleportation, the vital building block of a quantum repeater [1] is based on the distribution and measurements of entangled photons. Furthermore, quantum communication protocols such as blind quantum computing [2], distributed quantum computing, clock synchronization [3] and realization of device independent security [4] relies on the distribution of entanglement. Other types of quantum communication protocols notably quantum key distribution [5], quantum oblivious transfer and one-time programs [6, 7] can be implemented based on a prepare and measure scheme or on entanglement. In this case the use of entanglement, although technologically more challenging, offers additional benefits. Since the source of entanglement contains no active encoding, there is no information present at the source. It can therefore be placed in a completely untrusted environment, and even be operated by an adversary, under the assumption of course, that the receiver stations can be trusted and are secured against (side channel) attacks. In addition, the source is also a completely passive device, no active element is required to generate or encode information onto the photons. There is also no need for quantum random number generators which can limit the launch rate of single photons. With entanglements, very small and robust sources [8-10] producing pairs with GHz rates [11] can hence be fabricated. Finally, specific network topologies like point-to multipoint and other tree-like mapping can be achieved with less resources using entanglement.

    The production and distribution of bi-partite entanglement, where entanglement is shared between two photons as the simplest and hence most prevalent method, has been demonstrated over long distances via free-space [12] or fiber-optic channels [13]. However, a network architecture needs to be implemented if more than two users want to share quantum entanglement between them. In early implementations, broadband entanglement sources were sliced through adoption of dense wavelength division multiplexing (DWDM) technology [14-16]. Alternatively, spatial switching has been considered as a replacement of the wavelength dimension in combination with spectrally narrow entanglement sources [17]. Through the incorporation of active switching, dynamic reconfiguration becomes possible. As a natural extension of both, active switches in combination with broadband sources and WDM slicing were employed to direct entangled photon pairs to the users in questions [18, 19]. The reconfigurability of ring and tree networks that is gained through WDM demultiplexing and spatial switching has been investigated in [20, 21].

Those schemes allowed different spectral allocations but always with a fixed rate of entangled photons. Another approach, using only passive wavelength multiplexers, was currently demonstrated [22], but apart from an all passive and static setup, this method lacks both the flexibility to change the rate of entanglement as well as the allocated wavelengths.

In this work, we present an experimental entanglement and photon pair distribution scheme with very much enhanced functionalities than in previous realizations. Our scheme permits flexible distribution between not only network users but also the central office hosting the entanglement and photon pair sources, as well as an on-demand allocation of wavelength channels. We also demonstrate the dynamic bandwidth allocation of entangled or pair photons between privileged users in the network. These advanced routing solutions are experimentally verified using a realistic fiber network architecture and can easily be scaled up to a larger number of users without the need for introducing more optical components and hence loss in the distribution channel.

The paper is organized as follows. Section II introduces the concept behind the flexible allocation of quantum bandwidth that serves entanglement distribution. Section III describes the quantum sources that constitute the network assets. The experimental setup used for performance evaluation is detailed in Section IV. The experimental results on entanglement and photon pair distribution are discussed in Section V, before Section VI addresses the dynamic switching performance for the proposed network scenario. Finally, Section VII concludes the work.

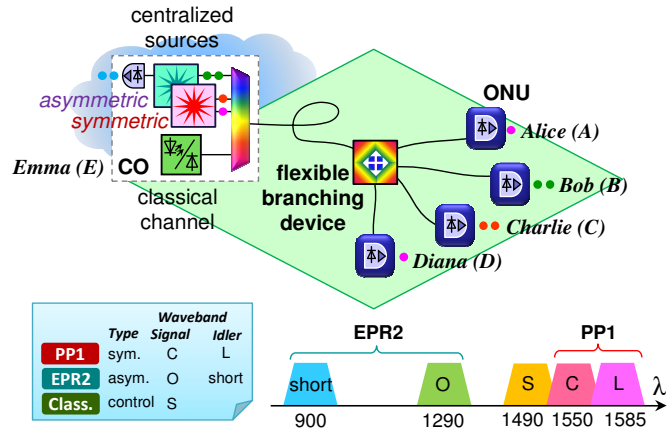

Fig. 1. Dynamic allocation of quantum bandwidth with a reconfigurable network node serves the flexible distribution of entanglement in an optical network.

## II. Flexible Entanglement / Photon Pair Distribution

Our network, as shown in Fig. 1, comprises of a central office (CO) which hosts two photon-pair sources (Emma **E**). These are connected via a feeder span to the branching node, which is comprised of an active space switch and several passive wavelength multiplexers to allow for the flexible entanglement or photon pair distribution to the four end-users (Alice **A**, Bob **B**, Charlie **C** and Diana **D**) at the optical network units (ONU), similar as we had introduced in brevity earlier [23]. These four users are connected to the branching node with shorter drop spans, similar as it applies to optical access networks. As a matter of fact, cloud-based radio access, where optical front/backhauling is collapsed over the optical wireline infrastructure, builds on such network topologies with a reach of up to 20 km. This reach value results from the stringent 5G latency requirements of 1 ms for a signal round trip, including both, bidirectional signal transmission and processing. The symmetric photon-pair source (PP1) at Emma distributes the photon pairs through the various mapping of the branching node to users **A** through **D**. Symmetric, in this case, implies that both photons are emitted at telecommunication wavelengths. PP1 produces two streams of photon pairs, with two wavelengths in the C- and another two wavelengths in the L-band (a photon pair always consists of a C/L-waveband combination). In order for Emma to share and communicate via entanglement to the other users, an Einstein-Podolsky-Rosen source (EPR2) is installed at the CO. This source has an asymmetric wavelength output, with one photon at 900 nm and the other in the O-band at 1290 nm. The short-wavelength photon is directly detected at **E**, as to provide the centralized hub with an entangled photon, too. The longer-wavelength photon is transmitted through the network to the users **A** trough **D**, according to the mapping of the branching node. Altogether, our network serves four users and one cloud service provider separated by optical fibers. To demonstrate the flexibility in allocating entanglement resources between those communication partners we chose 6 exemplary scenarios of the distribution map, as introduced in Fig. 2.

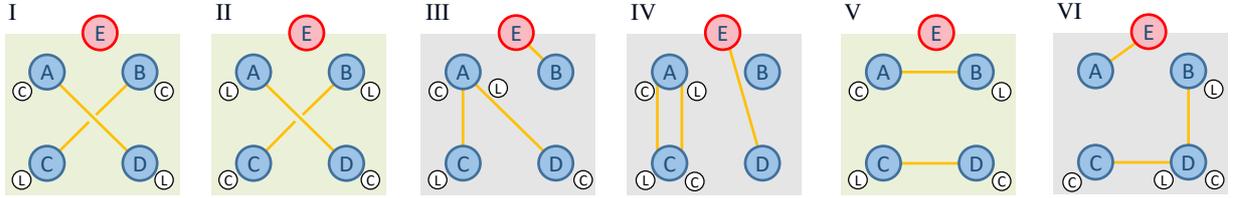

Fig. 2. Distribution map for user- and cloud-centric sharing of single and entangled photons in scenarios I to VI.

*Mapping I*: In this case Alice and Diana as well as Bob and Charlie share pair photons from PP1, whereby Alice and Bob receive photons in the C-band and Charlie and Diana the photons from the L-band. In this scenario Emma does not share entanglement at the same time. This first scenario is purely user-centric.

*Mapping II*: This setting is similar to mapping *I* in terms of user allocation (**A**-**D** and **B**-**C** share again pair photons), but this time the wavelength assignment is different, with **A** and **B** receiving the L-band photon and **C** and **D** the photons from the C-band.

*Mapping III*: In this scenario, Alice has a special role, she shares pair photons with two other users (**C** and **D**) simultaneously, in a kind of concentration scenario, as the rate of measured photons at **A** is elevated. **A** receives of course both the L- and C-band photons with the rest shared between **C** (L-band) and **D** (C-band). Since Bob is now free, he can use EPR2 to share its entanglement with Emma at the same time. The latter provides a cloud-centric perspective, with Bob being a premium user that enjoys a high-rate quantum pipe to the CO.

*Mapping IV*: Here we assign special status to Alice and Charlie, which now share both streams of pair photons from PP1, providing them a much higher bandwidth. Since neither **B** nor **D** participate in this exchange, we chose to share additional entanglement between **D** and **E**.

*Mapping V*: This setting demonstrates the different user allocation, as compared to mapping *I*, by distributing pair photons between **A**-**B** and **C**-**D**.

*Mapping VI*: Finally, we demonstrate a different wavelength and user allocation as seen in mapping *III*, with pair photons and entanglement being shared between **D**-**B**, **D**-**C** and **A**-**E**.

The switching node that acts as a flexible branching device and as an enabler for a reconfigurable distribution mechanism, is introducing optical losses in excess to the transmission losses of the feeder and drop fiber spans. Given this elevated loss budget, this work will investigate if a distribution of quantum resources can be in principle supported when employing state-of-the-art pair sources in this particular quantum overlay architecture. For example, while for the use-case of real-time encryption of classical data through one-time pad would require secure-key rates in the order of 10 Gb/s, which is currently touted as unfeasible even with dedicated quantum key distribution systems, a periodic key exchange for classical encryptors such as based on AES would already benefit from a small quantum-enabled secure-key rate as long as it is above zero for the given optical loss budget.

It shall also be stressed that the branching node, though it can be in principle co-located at the CO with Emma, is practically not controlled by any of the network users (**A**-**E**). In a realistic deployment it will be instead instructed by the network operator after receiving requests for resources by the end-users (**A**-**D**) or the cloud service provider (**E**).

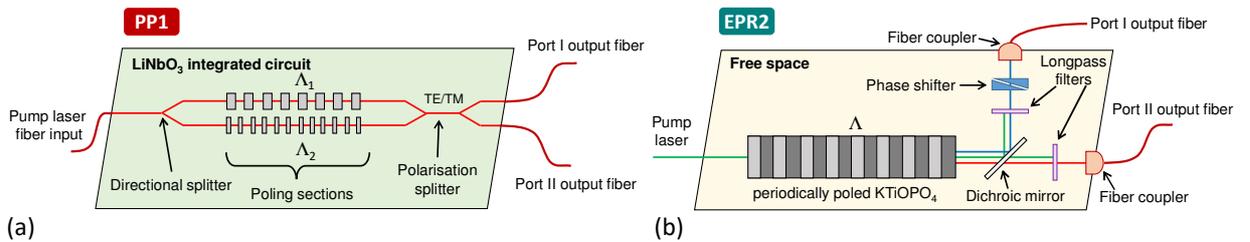

Fig. 3. Employed pair sources. (a) Symmetric source for emission in C- and L-band. (b) Asymmetric source for short-waveband and O-band emission.

## III. Photon-Pair Sources

Both sources, PP1 and EPR2, are based on spontaneous parametric down-conversion (SPDC) in periodically poled nonlinear media. Two sources are employed in this work, comprising a symmetric and an asymmetric source, as sketched in Fig. 3.

*A.    Symmetric photon-pair source*

The first source is integrated on a lithium niobate (LiNbO$_3$, LN) waveguide platform [24]. A continuous-wave fiber laser with wavelength $\lambda_p$ = 780 nm, pigtailed to the chip, pumps two periodically poled LN structures simultaneously, both supporting type-II phase-matching for photon-pairs symmetrically located below and above 2×$\lambda_p$. However, the two poling periods $\Lambda_1$ and $\Lambda_2$ differ in such a way as to produce photon pairs at different wavelength channels. Waveguide 1 with poling period $\Lambda_1$ will emit horizontally polarized photons in the C-band (~1540 nm) and vertically polarized L-band photons (~1590 nm), as shown in Fig. 4. On the other hand, waveguide 2 with period $\Lambda_2$ generates perpendicularly polarized photons at ~1510 and ~1610 nm. The photon pairs from both crystals are split by a polarization beamsplitter and fed to two output fibers. In principle the source can produce polarization entangled photon states [12] but we operate it as a non-entangled photon pair source to provide four different output wavelengths, two in the C- and two in the L-band.

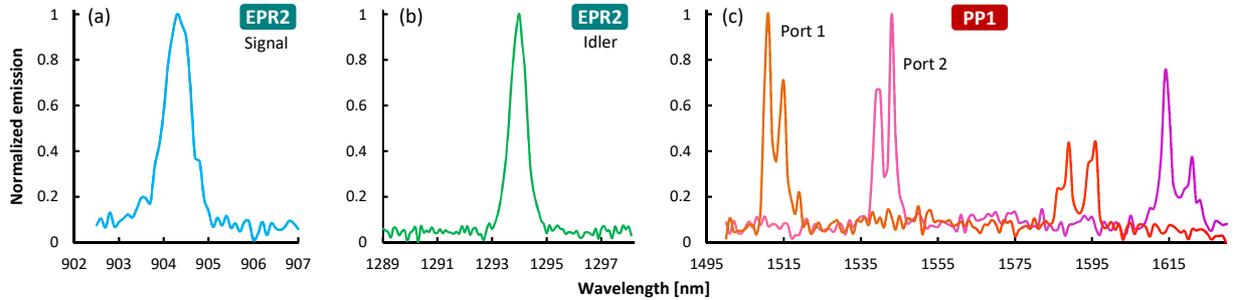

Fig. 4.  Emission spectra for both sources. (a) Short-waveband and (b) O-band emission of EPR2. (c) C- and L-band emission of PP1.

*B.    Asymmetric EPR source*

The second source exploits a recently discovered phenomenon that we call collinear double down-conversion [25, 26]. This effect is again based on the quasi-phase-matching technique where momentum conservation is satisfied by periodic alternation of the nonlinear coefficient along the nonlinear crystal. The length of the periodic poling domain $\Lambda$ depends on the absolute value of the momentum mismatch $\Delta k = k_p - k_s - k_i$, where $k_{p,s,i}$ is the pump, signal and idler momentum, respectively. Since one poling periodicity $\Lambda$ provides phase-matching for positive and negative $\Delta k$, one periodically poled nonlinear crystal, pumped by a single laser, can emit two different photon pairs at the same time: one with phase mismatch $\Delta k$ and one with $-\Delta k$. We use this property to satisfy phase-matching for two SPDC processes with same wavelengths but interchanged polarizations: $\lambda_{H+} = \lambda_{V-}$ and $\lambda_{V+} = \lambda_{H-}$ where $\lambda_{H+}$ ($\lambda_{V-}$) is the wavelength of the horizontally (vertically) polarized photon of the SPDC process with positive (negative) phase mismatch. Our method thereby allows for entanglement generation by unidirectionally pumping a single, uniformly poled nonlinear crystal, a setup with unparalleled simplicity. In our experimental implementation, a pump laser of wavelength $\lambda_p$ = 532 nm impinges a periodically poled KTiOPO$_4$ crystal (ppKTP), generating a signal beam in the near infrared ($\lambda_s$ = 904 nm) and an idler beam in the O-band ($\lambda_i$ = 1294 nm), both of them in horizontal and vertical polarization. While the signal beam is detected locally at the source using a silicon avalanche photo diode, the idler photons are fed into the optical network for entanglement distribution.

*C.    Emission spectra*

Figure 4 presents the normalized emission spectra of photon-pair and EPR sources. The asymmetric source (EPR2) features a spectrally narrow (~1 nm) signal and idler emission at the short-wavelength band and at the O-band, centered at 904 and 1294 nm, respectively. This wavelength pair fits well to the cloud-centric application scheme, where one photon of the pair is to be detected locally. Since no fiber transmission is required in this case, a high-efficiency silicon-based single-photon avalanche photodiodes (SPAD) can be employed for this purpose.

The emission spectrum of the symmetric source (PP1) is confined to the C- and L-band and can be therefore easily multiplexed to the O-band transmission wavelength of EPR2. The emission bandwidth for each of the peaks is ~6 nm. Since the emission at each of the output ports of PP1 occurs in both, the C- and the L-band, demultiplexing can be accomplished by C/L waveband splitters in order to tailor the delivered spectrum to the network users.

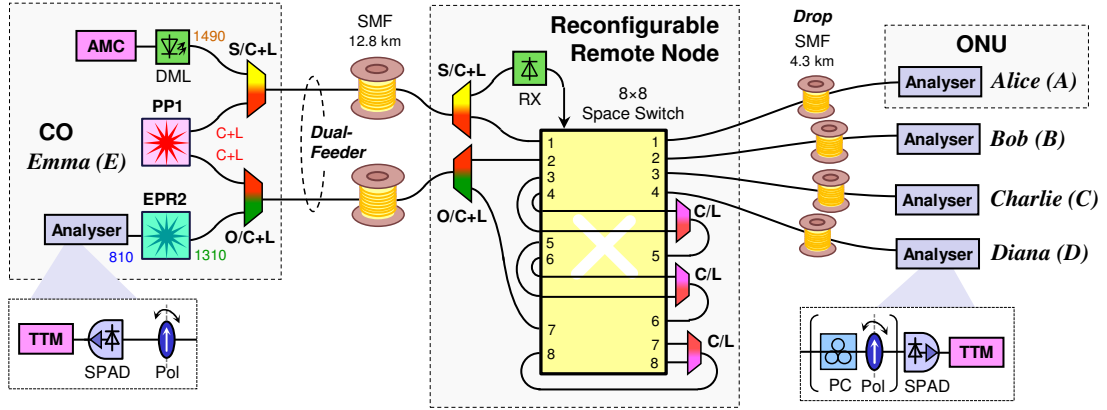

Fig. 5. Experimental setup for entanglement distribution with an architecture-on-demand node.

## IV. Experimental Setup and Node Characterisation

The experimental setup is presented in Fig. 5 and connects multiple network users in a tree-shaped network via a remote node to the CO. The CO hosts the symmetric (PP1) and the asymmetric (EPR2) photon-pair sources, and a classical channel that serves the auxiliary management and control (AMC) of the distribution network. The AMC channel is implemented through a directly modulated laser (DML) operating at 1490 nm, and a PIN-based photoreceiver at the remote node. The single-photon outputs of the sources are multiplexed with waveband combiners together with the AMC channel towards the optical distribution network. The fiber network consists of a dual-feeder and single-drop span and builds on ITU-T G.652B-compatible standard single mode fiber. The averaged kilometric fiber loss was 0.21 and 0.39 dB/km at 1550 and 1310 nm, respectively. According to the spectral allocation of the sources, each of the 12.8 km trunk fiber spans feeds the remote node with a C+L band signal/idler from the symmetric pair source, together with either the classical control channel in the S-band, or the O-band single photon output of the asymmetric EPR source. The dual feeder tree configuration is motivated by the brown-field layout of classical optical access networks, which foresee a passive power split in their optical distribution network. Such power splitting stages allow a 2×$N$ layout between two feeder fibers and $N$ drop segments for $N$ end-users, while it also supports resiliency features in its feeder segment to prevent failure in case of a feeder fiber cut [27]. At the quantum overlay, the two feeder fibers are advantageously applied to reduce the number of multiplexing stages that are required at the feeder segment, thus reducing the insertion loss.

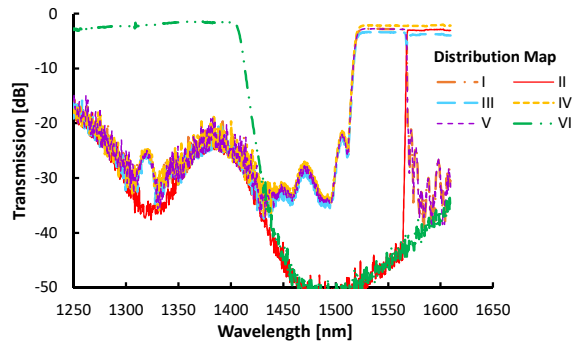

Fig. 6. Optical transmission spectra for the remote node. Results are shown for the entire distribution map of Alice' drop port.

At the remote node the signals are then flexibly distributed by the reconfigurable remote node according to the distribution map described earlier. For this purpose an 8×8 space switch in loopback-architecture is being instructed through the AMC channel. This classical control channel is only active during a reconfiguration, which ensures that no additional Raman noise is introduced [28] to the vulnerable quantum channels. While the four users are dedicated to half of the switch output ports, the remaining four ports are exploited in a foldback configuration that allows to synthesize a node architecture on demand [29]. Spectral tailoring of the quantum signals can be conducted through waveband demultiplexing and subsequent re-composition based on C/L waveband splitters. Moreover, the space switch allows direct routing of the O-band quantum channel to any of the users. The transmission windows of the reconfigurable node at Alice' interface are shown in Fig. 6 for the different distribution mappings. Distribution of the quantum signals in the three wavebands is realized at fiber-to-fiber insertion losses of 2.2 dB in the O-band, and 2.8 to 3.8 dB in the C- and L-band, depending on the actual loopback configurations used to compose the distribution of photon pairs. Rejection of the S-band in all settings ensures that the AMC channel is spectrally suppressed towards the users, in addition to its temporal multiplexing.

It shall be stressed that a reconfigurable node such as applied in the present work does not necessarily require a local power supply and can be fed by means of optical energy harvesting [30], thus retaining a fully-passive nature for the optical distribution network.

The optical network units (ONU) are connected through a 4.3 km drop-fiber span to the remote node and host an analyzer module to evaluate the transmission performance of the quantum signals.

User-centric detection of the quantum signals is conducted through InGaAs SPAD with a detection efficiency of 10% at 1550 nm and 12% at 1290 nm (**A**-**D**). Cloud-centric detection is facilitated by a Si SPAD with a detection efficiency of 5% at 900 nm (**E**). For the pairwise coincidence measurements between the **A** to **D** combinations, two InGaAs detectors were used. While one of them operated in free-running mode, the other InGaAs operated in a quasi free-running mode by selecting the fastest possible gate frequency (1 MHz) and a large gate window of 100 ns, resulting in an apparent efficiency of ~1% (as the combination of duty cycle and detection efficiency). The dark count rates for the free running InGaAs SPADs was 520 c/s with 30 μs deadtime, and 670 c/s for the detector in quasi free-running mode with a dead time of 10 μs. The dark count rate for the Si SPAD was 110 c/s (**E**). Registration of the detector events is performed through a time-tagging module (TTM), which was then used to calculate coincidence-count rates between the various users. A state analysis was performed to evaluate the performance of entanglement distribution, for which a polarization analyzer unit has been included at the CO and at the ONUs. Six different optical-path configurations have been evaluated, as summarized in Fig. 2.

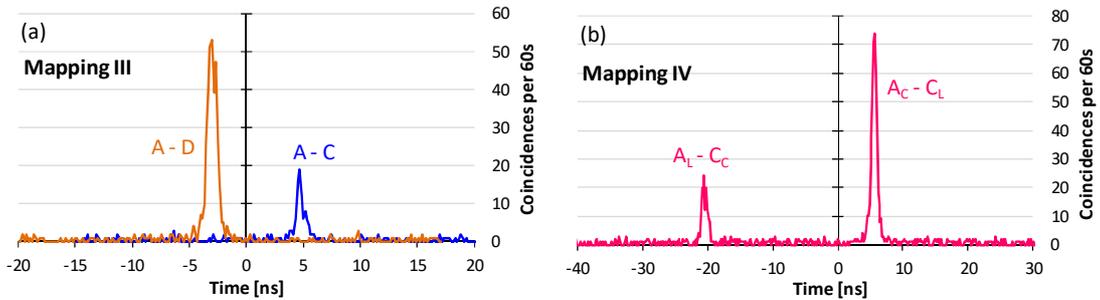

Fig. 7. Time-correlation measurements for mappings III and IV. The plots depict the relative time delay between detection events at two user sites. Peaks in the plots indicate simultaneously generated (and therefore correlated) photon pairs. (a) Alice shares photon pairs with Charlie and Diana at the same time. (b) Alice and Charlie share photons from both SPDC pairs generated by the symmetric source PP1.

**V. Static Distribution Mapping. Results and Discussion.**

The performance for single and entangled photon distribution was evaluated for various mappings in static operation. Tables I and II summarize the main findings for the implemented mappings in the network.

*A.      Distribution of single photons*

For the distribution of photon pairs from PP1 we measured the coincidence rates always between two user pairs. The coincidence counts were averaged for each measurement over 60 seconds with the average value and standard deviation (1σ) listed in Table I. Average coincidence rates vary between 2 and 6 coincidence counts per second (cc/s). There is also a clear asymmetry between the two streams of pair photons, very notable in mappings I and II and III. The difference in the measured rate is however consistent with the mapping of the two streams (a certain C/L-band pair has lower rate) and we ascribe the difference to a lower generation amplitude as seen in the spectral distribution in Fig. 4 for PP1. Pair concentration was successfully achieved in mappings II and IV as indicated by the higher rate for the users **A** and **D**. The increased bandwidth of the coincidence rate between users **A** and **C** was observed in mapping IV, where the total detected rate was more than 16 cc/s. The corresponding coincidence measurements between signal/idler photons are presented in Fig. 7. The fact that this is twice as much as the combined rates observed in the other scenarios can be explained by lower losses in the remote node which is only traversed once in this configuration.

We would like to stress that the low rates are not impaired by the flexible branching node but by the performance of the employed detectors and the imperfection of the employed photonic integrated PP1 source, which was not amenable to further tuning.

Table I
Coincidence Rates for Single-Photon Distribution (PP1)

| Mapping (User-Pairs) | Users / Performance | Pair 1 Coincidences | Pair 2 Coincidences |
|---|---|---|---|
| **I** A-D, B-C | mean | 5.9 | 1.2 |
| | std. dev. | 1.2 | 0.5 |
| **II** A-D, B-C | mean | 1.4 | 5.9 |
| | std. dev. | 0.7 | 1.5 |
| **III** A-C, A-D | mean | 1.2 | 4.7 |
| | std. dev. | 0.6 | 1.1 |
| **IV** $A_C$-$C_L$, $A_L$-$C_C$ | mean | 8.8 | 7.8 |
| | std. dev. | 1.4 | 1.2 |
| **V** A-B, C-D | mean | 5.8 | 2.3 |
| | std. dev. | 1.2 | 0.7 |
| **VI** B-D, C-D | mean | 2.6 | 5.8 |
| | std. dev. | 0.8 | 1.0 |

Values are provided per second.

### B.    Distribution of entangled photons

The polarisation entanglement shared out by EPR2 between **E** and the other users was also analysed together with the detected coincidence rate of entangled photons. Count rates were again averaged for 60 seconds. To characterise the entanglement, a visibility measurement was performed by placing polarisers in front of the detectors at **E** and the other user and measuring the correlations for the following eight polariser settings: HH, HV, VH, VV, AD, AA, DA, and DD, where H indicates horizontal, V vertical, D diagonal (45°) and A anti-diagonal (-45°) polarisation.

Polarization entanglement sourced by EPR2 was measured in a back-to-back configuration with 190.7 cc/s ($1\sigma$ = 13.2 cc/s) and an entanglement visibility of $V = 0.955 \pm 0.033$. For the three mappings III, IV and VI, where entanglement is shared with Emma, the measured visibility is listed in Table II. As one can see, there is a drop in the entanglement to about 0.85 from the back-to-back measurement. The reason for this is the low count rate which makes it difficult to perform a correct polarization compensation of the rotation induced by the long fibers. If the birefringence is not compensated exactly, the visibility is reduced. Nevertheless, the visibility measurements clearly exceeded the classical limit of $1/\sqrt{2} \approx 0.707$. At a pump power of 8.5 mW for EPR2, the local-to-remote brightness (**E-X**) amounted to $29.3 \pm 2.3$ cc/s of polarization-entangled photons. This value agrees well with the expected rate due to transmission loss at feeder fiber, branching node and drop fiber.

Table II
Visibility for Entangled-Photon Distribution (EPR2)

| Mapping (User-Pairs) | Users / Performance | Visibility |
|---|---|---|
| **III** E-B | mean | 0.841 |
|  | std. dev. | 0.075 |
| **IV** D-E | mean | 0.886 |
|  | std. dev. | 0.067 |
| **VI** A-E | mean | 0.842 |
|  | std. dev. | 0.109 |

## VI. Dynamic Allocation of Quantum Bandwidth

Finally, dynamic switching of available quantum bandwidth has been conducted by instructing the remote node to reconfigure its matrix. The entire distribution map of Fig. 2 has been cycled through, with each of its six configurations being applied for the time duration of multiple seconds. Four InGaAs SPADs have been employed for this measurement in order to register the delivered photon rates. The dark count rates were 520, 1300, 1525 and 1845 c/s (**A**-**D**).

Figure 8(a) shows the registered count rates at all network users, when implementing an electrical AMC channel that bypasses the optical feeder segment. In addition, Fig. 8(b) shows the same dynamic bandwidth allocation for an optically implemented AMC channel in the S-band, which stands in co-existence with the quantum channels that are transmitted over the same feeder fiber.

The impact of a change in the spectral allocation when reconfiguring the node is directly visible in the registered photon counts at the remote users (Alice to Diana), while the centralized detector (Emma) keeps a steady rate independent of the node configuration for the optical network. The actual gain in delivered rate that results from the reconfiguration of the branching node depends not only on the obtained spectral tailoring, but also on the optical path within the branching node. This path-dependent insertion loss of the node has been discussed earlier in Fig. 6 for Alice' port, whereas the spatial switch shows a rather good loss uniformity with a peak-to-peak deviation of 0.23 dB among its ports. A doubling of spectral bandwidth, as for example supported through mappings III and IV in case of Alice, does therefore not lead to a doubling in rate since the multiplexer losses partially erode the rate gain. Still, these mappings lead to the highest rates obtained. Mappings that provide a lower rate, such as II in case of Alice, do not mean that photons are artificially suppressed but rather distributed to other users. This allows flexible bandwidth provisioning within the network. With reference to the lowest configurable output rate at the C/L-band distributed photons of the user-centric distribution, the delivered photon rate for Alice can be dynamically adjusted by a factor of 3.2 or 5.1 dB after subtraction of detector dark counts.

As an alternative to the branching node, a spatial 1×4 switch could be inserted in order to decrease the optical loss of the distribution node. However, this would prevent the simultaneous provision of pair photons to more than one user, which might pose a serious limitation when the number of users scales up. Broadcasting of pair photons with a 1×4 power splitter would introduce a large optical loss of ~6.4 dB, which is much higher than that of the reconfigurable branching node (Fig. 6).

The fast switching time of the space switch of the node does not cause a gap larger than the 50-ms resolution for the acquisition of the delivered photon rates. With this fast switching, a quasi-hitless distribution of quantum signals is accomplished.

Moreover, when implementing the AMC channel in the optical domain at 1490 nm, there is no impact observed at the switching transients. This is explained by the short 1.6 ms duration for the transmission of a reconfiguration word, the temporal gating and its spectral filtering at the quantum channels. This proves the feasibility of a co-existing classical channel that is dedicated to management of the network assets.

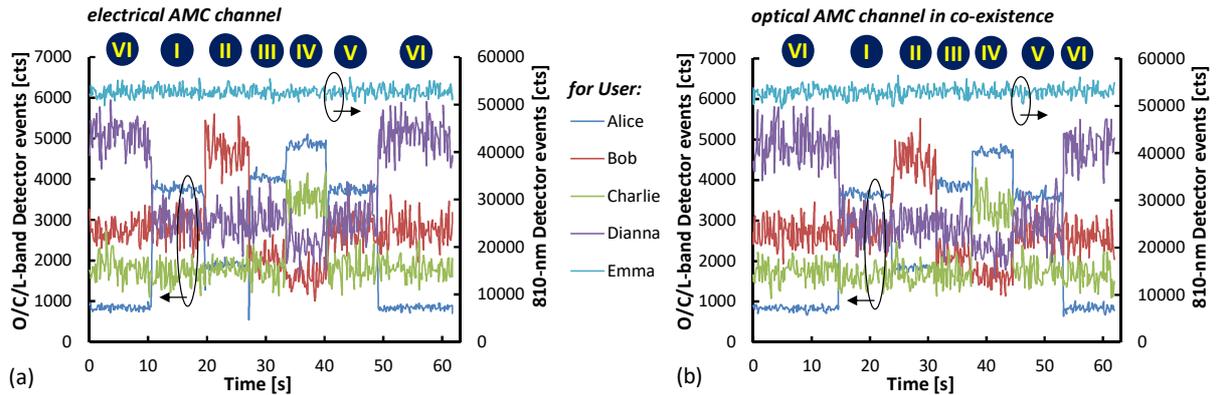

Fig. 8. Dynamic switching of the remote node configuration to implement the entire distribution map with its configurations I to VI. Results are shown as detector events for all users in case of having an AMC channel implemented in (a) the electrical and (b) the optical domain.

## VII. Conclusion

We have experimentally demonstrated the flexible provisioning of single and entangled photons among multiple network users. Dynamic allocation of quantum bandwidth between end-nodes is accomplished through means of reconfigurable spectral slicing from the O- to the L-band and subsequent spectral tailoring in combination with broadband EPR sources and optical space switching. Two representative network schemes have been investigated, including an asymmetric, cloud-centric scenario where a high-rate entanglement pipe is provided to a centralized location, and a user-centric scenario where photon pairs are distributed to two end-nodes, without connecting the central office. Transmission results in a 1:4 split, 17-km reach optical tree network with reconfigurable remote node have indicated that the rate of delivered photons can be adaptively adjusted in six representative photon distribution mappings. Moreover, the integration of a classical control channel within the quantum network has been found not to result in a penalty for the distribution of single and entangled photons. By leveraging optical switching technology, agile and hitless reconfiguration is obtained at the quantum plane. Improvement in terms of delivered rates is expected for higher brightness of the pair sources in combination with high-end detector technology. In future all InGaAs detectors should be free running, increasing the detection rate by a factor of 10. If higher efficiency detectors such as superconducting single-photon detectors are used, an increase in the detection rate of up to three orders of magnitude could be achieved. Moreover, to demonstrate the full potential of our routing scheme, a broadband entanglement source needs to be installed instead of the PP1 source. In this way one can achieve entanglement distribution between all users and also an increase in number of users by exploiting a larger optical emission bandwidth.

## VIII. Acknowledgement

The authors would like to thank the Integrated Quantum Optics group at the University of Paderborn for providing the entangled photon source on chip. This work has received funding from the EU Horizon-2020 R&I programme (grant agreement No 820474).